\documentclass[11pt,twoside]{article}
\usepackage{asp2010}

\usepackage{amsmath}
\usepackage{float}

\newcommand{\vel}{\ensuremath{\varv}}

\newcommand{\vinfty}{\ensuremath{\vel_\infty}}
\newcommand{\vcl}{\ensuremath{V_\mathrm{cl}}}
\newcommand{\veldis}{\ensuremath{\vel_\mathrm{dis}}}
\newcommand{\rcl}{\ensuremath{r_\mathrm{cl}}}
\newcommand{\kms}{\ensuremath{\mathrm{km}\,\mathrm{s}^{-1}}}
\newcommand{\Halpha}{\ensuremath{\mathrm{H}\alpha}}

\begin{document}

\title{3-D Monte Carlo radiative transfer calculation of 
resonance line formation in the inhomogeneous expanding stellar wind}

\author{Brankica \v{S}urlan$^{2,1}$, Wolf-Rainer Hamann$^3$, Ji\v{r}\'{\i} 
Kub\'at$^1$, Lidia Oskinova$^3$, Achim Feldmeier$^3$}

\affil{$^1$Astronomick\'y \'ustav AV \v{C}R, Fri\v{c}ova 298, 251 65
Ond\v{r}ejov, Czech Republic}

\affil{$^2$Matematicko-fyzik\'aln\'{\i} fakulta UK, Ke Karlovu 3, 121 16
Praha 2}

\affil{$^3$Institut f\"{u}r Physik und Astronomie, Universit\"{a}tsstandort 
Golm, Karl-Liebknecht-Str. 24/25, Potsdam, Germany}

\begin{abstract}

We study the effects of optically thick clumps, non-void inter-clump medium, 
variation of the onset of clumping, and velocity dispersion inside clumps on 
the formation of resonance lines. For this purpose we developed a full 
3-D Monte Carlo Radiative Transfer (MCRT) code that is able to handle 3-D 
shapes of clumps and arbitrary 3-D velocity fields. The method we developed 
allows us to take into account contributions from density and velocity wind 
inhomogeneities to the total opacity very precisely. The first comparison
with observation shows that 3-D density and velocity wind inhomogeneities
have a very strong influence on the resonance line formation, and that they 
have to be accounted for in order to obtain reliable mass-loss rate 
determinations.
\end{abstract}

\section{Introduction}

There is a lot of observational and theoretical evidence
that the winds of hot stars are inhomogenoues. Stochastic variable 
structures in the \ion{He}{\sc ii} $\lambda$ $4686 \AA$ emission line 
in $\zeta$ Pup were revealed by \citet{1998ApJ...494..799E}, and 
explained as an excess emission from the wind clumps.
\citet{2005A&A...440.1133M} investigated the {\Halpha} line-profile 
variability for a large sample of O-type supergiants. They concluded 
that the properties of this variability can be explained by a wind model 
consisting of clumps.

Understanding of clumping is a prerequisite to obtain correct
mass-loss rates from modeling the stellar spectra
\citep[see][]{2006ApJ...637.1025F, 2005A&A...438..301B,
2006A&A...454..625P, 2007A&A...476.1331O}.

A possible way how to study wind clumping is to use simplifications. 
The approximate treatment of clumping is usually based on the ``microclumping'' 
(clumps are optically thin at all frequencies) or ``macroclumping'' 
(clumps may be optically thick for certain wavelengths) approaches. 
Common additional assumptions are the void interclumped medium (ICM) 
and the monotonic velocity field. The enhancement of the density inside 
clumps is usually described by the so called ``clumping factor'' $D$.

In the most recent studies of \citet{2010A&A...510A..11S, 2011A&A...528A..64S} 
it was shown that the detailed density structure, non-monotonic velocity field, 
and the ICM are all important for the line formation in the massive hot star 
winds. In order to be able to study the 3-D nature of wind clumping and to take
into account all properties of wind clumping mentioned here, we developed a
realistic full 3-D MCRT code (\v{S}urlan et al., in preparation). Here we
present the basic effects of the different clump properties on the resonance 
line formation and present the first preliminary comparison with observation.

\section{The clumped wind model}

Our wind model is grid-less and does not require any symmetry. Instead of 
using a predefined grid, we introduce an adaptive integration step $\Delta r$. 
First we create a snapshot of clumps inside the wind and then we follow  
photons along their paths using the Monte Carlo approach. The photon 
frequencies are expressed in the local co-moving frame. All distances
are expressed in units of the stellar radius ($R_{\ast}$).

The velocity field can be arbitrary in our wind model. However, here we 
assumed a radial monotonic velocity $\vel_{r}$ field and the standard 
$\beta$-velocity law for the underlying smooth wind ($\vel_{r}=\vel_{\beta}$).  
The velocity inside clumps $\vel_{c}$ is expressed as 
\begin{equation}
\label{veldev}
\vel_{c}(r)=\vel_{\beta}(r^\mathrm{c}_{i}) - \veldis(r)\,\frac{r-r^\mathrm{c}_{i}}{l_{i}},
\end{equation}
where $\veldis(r)=m\,\vel_{\beta}(r)$ describes the velocity dispersion
using the deviation parameter $m$ ($ 0 < m\le 1$). The monotonic velocity at 
the position $r^\mathrm{c}_{i}$ is $\vel_{\beta}(r^\mathrm{c}_{i})$, 
and $l_{i}$ and $r^\mathrm{c}_{i}$ are the radius and the absolute 
position of the center of the $i$-th clump, respectively. 

The opacity of the wind is calculated using the parameterization given by 
\cite{1980A&A....84..342H},

\begin{equation}
\chi(r)=\frac{\chi_{0}}{r^{2}\frac{\vel_{r}}{\vinfty}}\, \phi_{x}; 
\quad \phi_{x}=\frac{1}{\sqrt{\pi}}\,e^{-x^{2}},
\label{opa}
\end{equation}
where $\chi_{0}$ is the opacity parameter that corresponds to the line strength, 
$\vinfty$ is the terminal velocity of the wind, and $\phi_{x}$ is the absorption
profile, which depends on the dimensionless frequency
$x=\left ({\nu}/{\nu_{0}}-1 \right) {c}/{\vel_{D}}$,
where $\vel_{D}$ is the Doppler-broadening velocity which comprises both 
the thermal broadening and microturbulence. The ionisation is assumed to be 
constant. 

For simplicity we assume a spherical shape of the clumps with a radius that  
varies with the distance $r$ from the star, $l=l(r)$. A clump
radius is determined using the clump separation parameter $L_{0}$ as
\begin{equation}
l(r)=l_{0}\,\sqrt[3]{r^{2}\,\varw(r)}; \quad
\text{where} \quad
l_{0}=L_{0}\sqrt[3]{\frac{3}{4\pi D}},
\label{l01}
\end{equation}
$\varw(r)=\vel_{r}/\vinfty$ is the velocity in the units of the terminal
speed $\vinfty$, and $D$ is the clumping factor. The average clump
separation $L$ is also depth dependent and given as
$L(r)=L_{0}\sqrt[3]{r^{2}\,\varw(r)}$.
The density inside clumps is assumed to be by the clumping factor $D$ 
higher than the smooth wind density. In the first approximation, $D$ 
is assumed to be independent on $r$ and given by
\begin{equation}
D=\frac{L^{3}(r)}{\frac{4\,\pi}{3}l^{3}(r)},\quad D\ge 1.
\label{D}
\end{equation}

The ICM density is determined by the interclump density parameter $d$ which is 
also assumed to be depth independent and $0\le d < 1$. If the factor $d=0$,  
the ICM is void, otherwise the ICM is rarefied by the factor $d$ 
compared to the smooth wind density. For the case of dense clumps and void ICM, 
all the mass of the wind is put into clumps, and the volume filling factor is
$f_{V}=1/D$. For the case of dense clumps and non-void ICM, the mass of the 
wind is distributed between clumps and ICM and it follows that 
$f_{V}=(1-d)/(D-1)$.

The total number of the clumps is determined by the condition of the mass 
conservation inside the wind. Clumps are created one by one and when the 
total volume of created clumps ($\vcl$) is larger than the total 
volume of the wind ($V_{w}$) multiplied by $f_{V}$, i.e. if 
$\vcl\ge f_V V_{w}$, then the number of created clumps represents the 
total number of clumps in our snapshot clump distribution.

\section{Effect of different clump properties}

\begin{figure*}[t]
\includegraphics[width=0.5\textwidth]{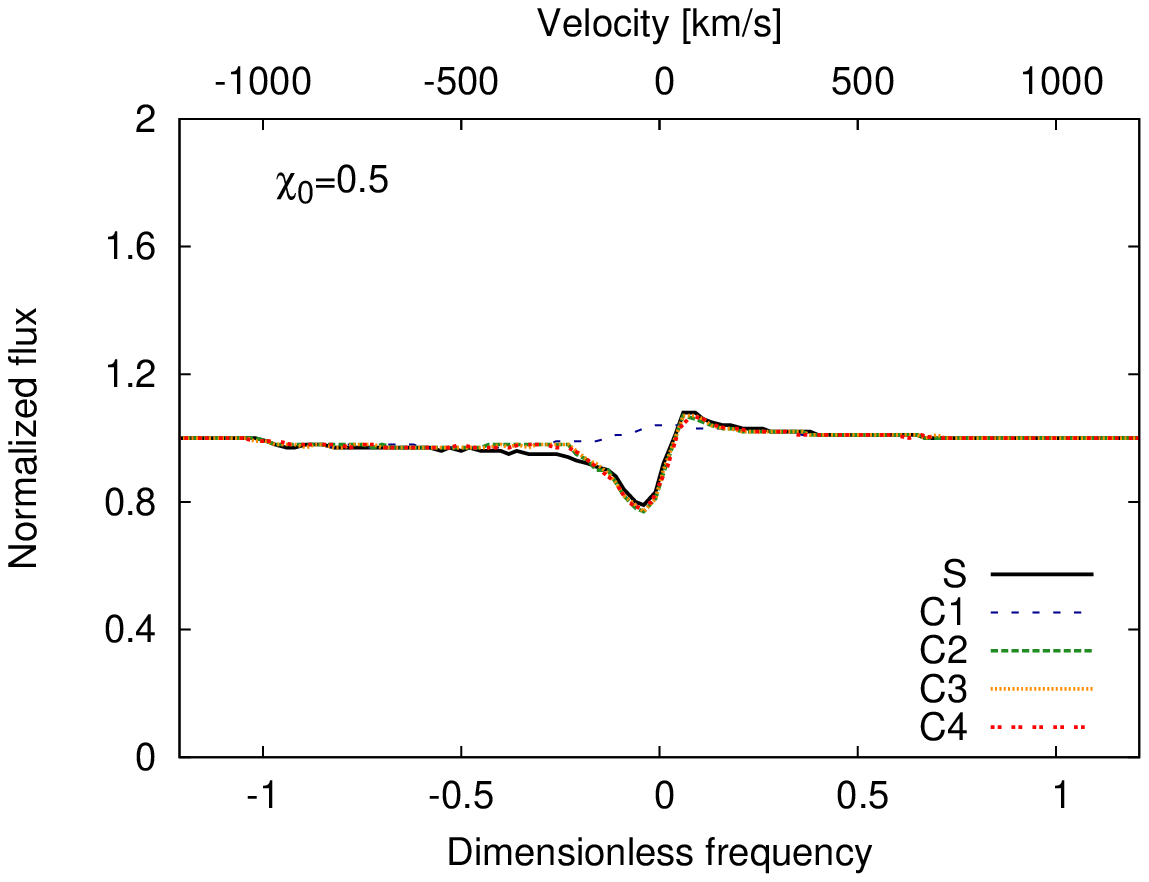}
\includegraphics[width=0.35\textwidth]{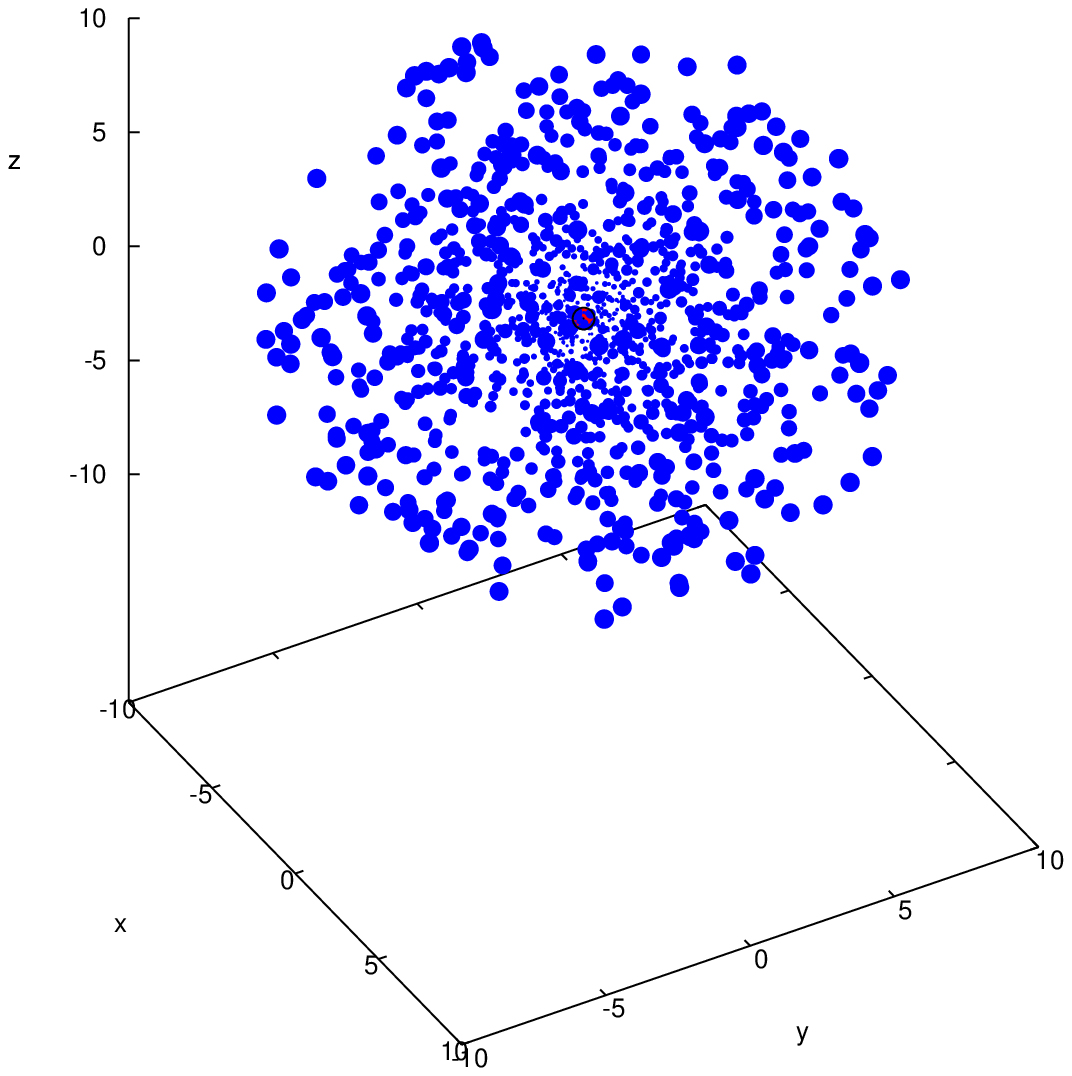}
\includegraphics[width=0.5\textwidth]{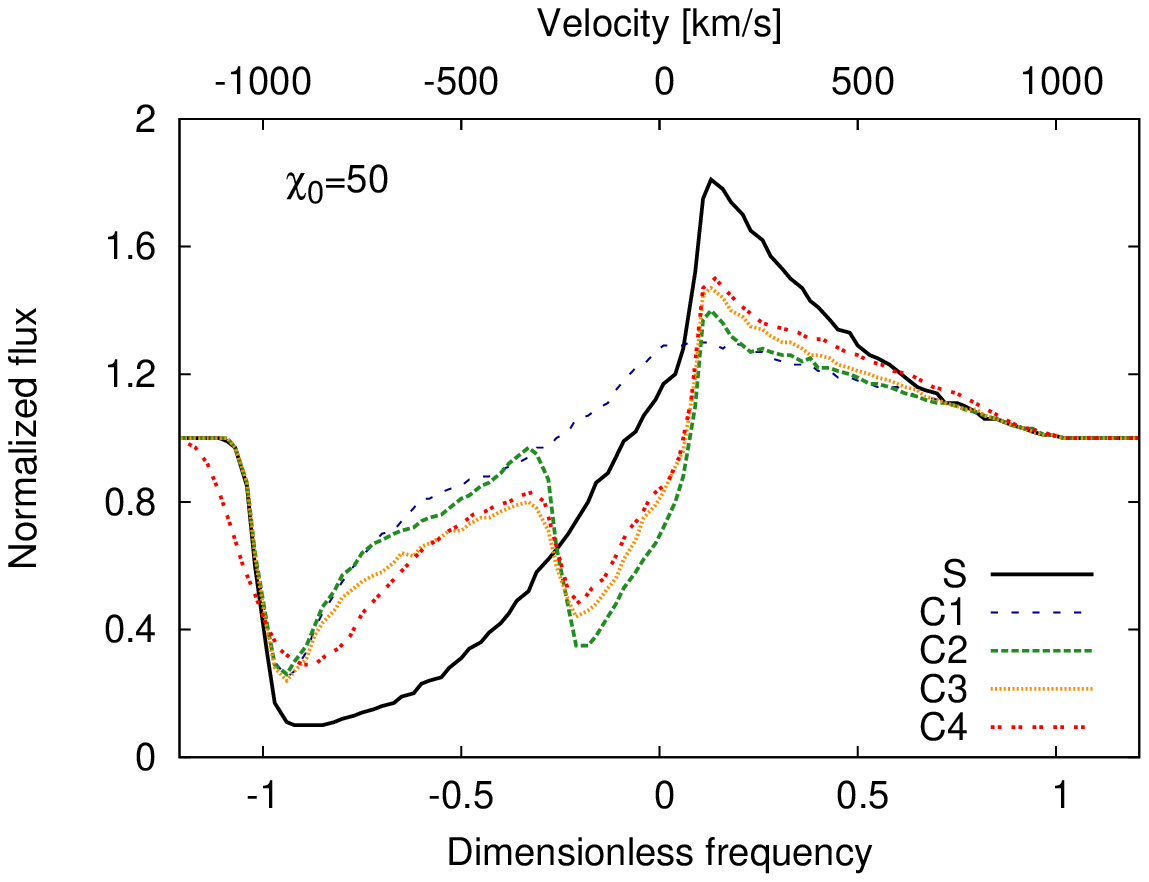}
\includegraphics[width=0.49\textwidth]{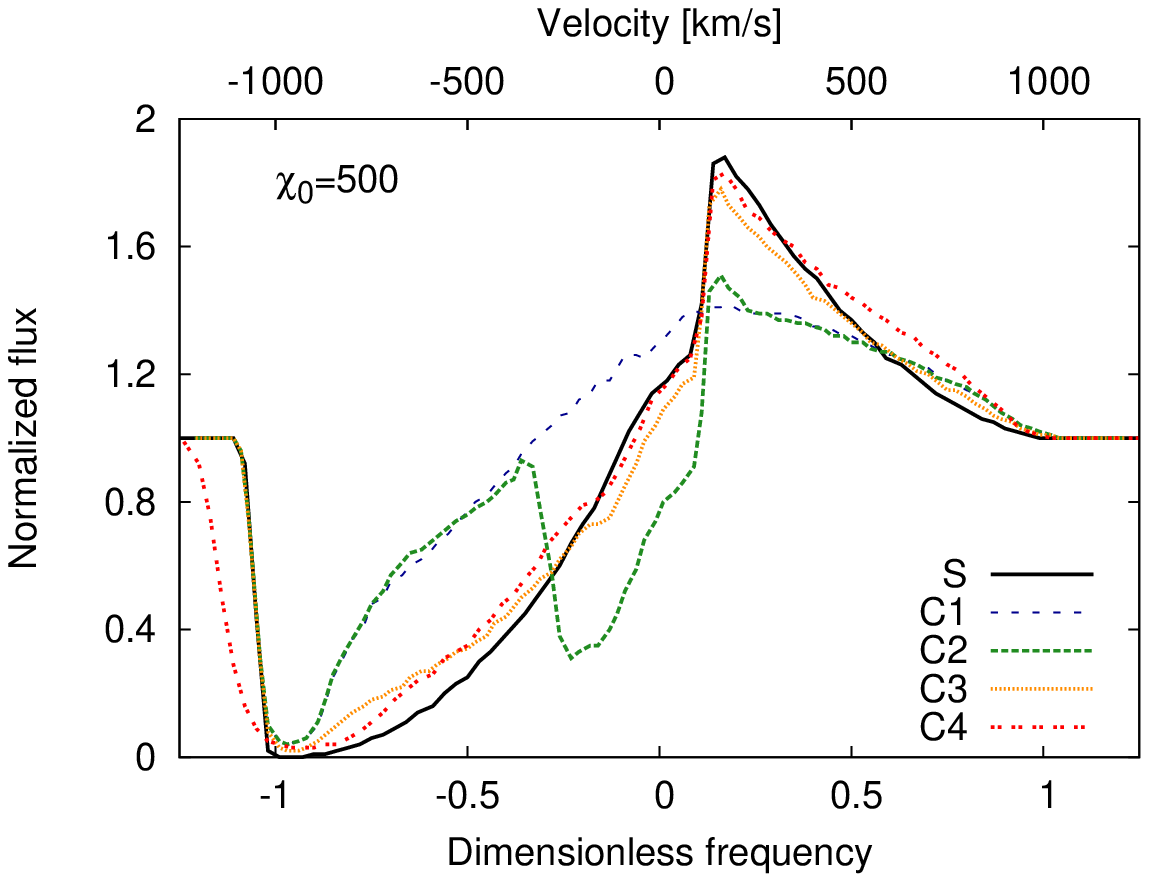}
\caption{Line profiles for weak (upper left panel), medium (lower left
panel), and strong (lower right panel) lines for the clump distribution
in the upper right panel. S -- the smooth model; C$n$ denotes the clumped 
models -- C1 ($\rcl=1$), C2 ($\rcl=1.3$), C3 ($\rcl=1.3$, $d=0.05$),  
C4 ($\rcl=1.3$, $d=0.05$, $\veldis=0.2\,\vel_{\beta}$).}
\label{fig:profiles}
\end{figure*}

By varying the model parameters we examined how macroclumping (defined by $L_{0}$ 
and $D$), non monotonic velocity field ($\veldis$), non void ICM ($d$), 
and onset of clumping ($\rcl$) may influence weak ($\chi_{0}=0.5$), 
intermediate ($\chi_{0}=50$), and strong ($\chi_{0}=500$) resonance line 
profiles. Calculations are performed assuming $\sim 10^{4}$ clumps inside wind,
which corresponds to $L_{0}=0.5$ for the wind with $r=10$. The clumping factor  
is set to $D=10$. The Doppler-broadening velocity is $\vel_{D}=50 \, \kms$. 
A snapshot of distribution of the clumps is shown in Fig.~\ref{fig:profiles}, 
upper right panel.

The first set of line profiles (C1 in Fig.~\ref{fig:profiles}) is 
calculated assuming that clumping
starts at the surface of the star ($\rcl=1$) with void ICM ($d=0$), 
and monotonic velocity field ($\veldis=\vel_{\beta}$). The weak (upper 
left panel), intermediate (lower left panel), and strong (lower right
panel) lines show strong line strength reduction compared to the smooth 
wind (full lines). If the wind is clumped,
more photons can escape through ``holes'' between clumps. This
leads to lower effective opacities, and, consequently, less absorption. The line 
strength reduction depends on the number of clumps in the wind. 
The absorption dip close to $\vinfty$ is caused by lower probability
of photon escape through lower number of velocity ``holes''.

In order to investigate how the onset of clumping influences line profiles, 
we calculated the second set of line profiles (C2 in Fig.~\ref{fig:profiles})  
assuming that clumping is only above $\rcl=1.3$. Other model parameters are
the same as for the first set of line profiles calculations. A very strong 
and broad absorption near the line centers appears. It is due to a higher 
absorption of the smooth part of the wind, which is in the region 
$1<r<\rcl$. Such deep absorption is not observed and it disappears 
if clumping starts closer to the wind base (perhaps even from the 
surface of the star) or if the ICM is sufficiently dense. The weak line 
almost reproduces the smooth wind model in this case. For a small value of 
$\chi_{0}$, the clumps are optically thin and they have a very small influence
on the line profile compared to the smooth wind.

Varying the ICM density parameter $d$, a significant impact on the line 
profiles was found, especially on strong lines. Keeping the same parameters as 
for the second set of line profile calculation, but choosing $d=0.05$ we  
calculated the third set of line profiles (C3 in 
Fig.~\ref{fig:profiles}). While the weak line is unaffected, the intermediate 
and strong ones show an increase of absorption in the blue edge of the line and
a decrease of absorption near the centre of the line. The degree of the line 
saturation depends on the ICM density. For a particular value of $d$,
saturation of strong lines and desaturation of the intermediate lines
can be achieved.

Taking into account the non-monotonic velocity field, an absorption 
at velocities higher than $\vinfty$ appears. Setting $\veldis=0.2\,\vel_{\beta}$ 
and keeping other model parameters the same as for the third 
set of line profiles, we calculated the fourth set of line profiles 
(C4 in Fig.~\ref{fig:profiles}). The higher $\veldis$, the larger
is the velocity span of clumps, and, consequently, there are more velocity
overlaps (the ``holes'' in the velocity field are smaller) and the probability 
of the photon escape is lower.

\section{Comparison with observations}

\begin{figure}[t]
\begin{center}
\includegraphics[width=1.\textwidth]{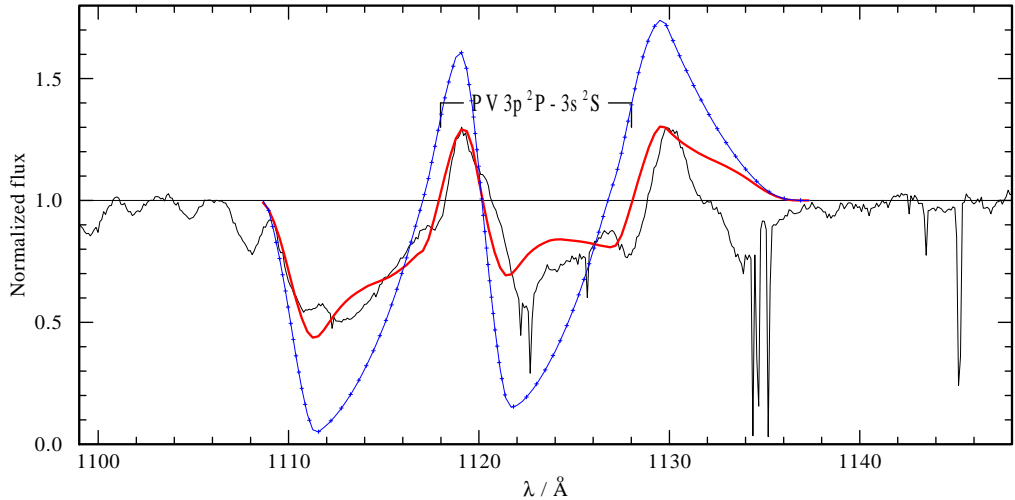}
\caption{Comparison of line profiles calculated using different wind
models with {\sc copernicus} observations (thin full line).The unclumped wind
model is shown by the full line with crosses, while  clumped model is shown
by the thick full line.}
\label{fig:comparison}
\end{center}
\end{figure}

As the first attempt to compare with observations, we try to fit the
\ion{P}{v} doublet of $\zeta$~Pup observed by the {\sc copernicus}
satellite (thin line in Fig.~\ref{fig:comparison}). As the terminal 
velocity of $\zeta$~Pup we adopt $\vinfty=2250\, \kms$, and
the value of $\beta=0.9$. The synthetic spectrum of the smooth and 
clumped wind models are calculated. The model  parameters are chosen 
to fit the observed spectrum best. It can be seen that the predicted 
unclumped (smooth) P-Cygni profile (full line with crosses) of \ion{P}{v}
is much stronger than the observed one (thin full line). However, 
the synthetic spectrum for the clumped model (thick full line)
fits the strength of the observed line very well.  
Therefore, using the unclumped model can lead to underestimating
the empirical mass-loss rates. These results are consistent with 
\cite{2007A&A...476.1331O} and
\cite{2010A&A...510A..11S,2011A&A...528A..64S}.

\section{Conclusions}

Using our 3-D Monte Carlo Radiative Transfer code, we investigate the 
influence of the 3-D stellar wind clumping on the resonance line 
formation. Our results show that clumping may significantly reduce the 
strength of lines and, if not included, may lead to a wrong 
mass-loss rate determination. The 3-D wind nature, the onset of clumping, non
void ICM, and non-monotonic velocity field are all important model
ingredients. Our models allow to obtain these parameters from the comparison
between model and observed lines, and thus gain deep insight in the
structure of stellar winds, and obtain reliable mass-loss rate measurements.

\acknowledgements 

This research was supported by 
grants GA~\v{C}R 205/08/H005, GA~\v{C}R 205/08/0003, and GA~UK 424411.

\bibliographystyle{asp2010}	
\bibliography{surlan}

\end{document}